\documentclass[pra,twocolumn,showpacs,superscriptaddress,floatfix,nofootinbib]{revtex4}
\usepackage{color}
\usepackage{hyperref}  
\usepackage{graphicx}
\usepackage{amsfonts}
\usepackage{footmisc}

     \let\d=\delta 
     \let\th=\theta  
\let\m=\mu        \let\x=\xi         \let\r=\rho
\let\s=\sigma \let\t=\tau    
     
 \let\D=\Delta

\font\tenmib=cmmib10\font\sevenmib=cmmib7\font\fivemib=cmmib5%
\mathchardef\Bl   = "0515  

\def\Bl   {{\mbox{\boldmath$ \lambda$}}}

\def\AA{{\mathcal A}}\def\FF{{\mathcal F}}
\def\DD{{\mathcal D}}
\def\BB{{\mathcal B}}\def\NN{{\mathcal N}}
\def\PP{{\mathcal P}}\def\II{{\mathcal I}}

\def\T#1{{#1_{\kern-3pt\lower7pt\hbox{$\widetilde{}$}}\kern3pt}}

\def\ie{{\it i.e.\ }}
\def\dpr{{\partial}}

\def\otto{\,{\kern-1.truept\leftarrow\kern-5.truept\to\kern-1.truept}\,}

\newdimen\xshift \newdimen\xwidth \newdimen\yshift \newdimen\ywidth

\def\ins#1#2#3{\vbox to0pt{\kern-#2pt\hbox{\kern#1pt #3}\vss}\nointerlineskip}

\def\eqfig#1#2#3#4#5{
\par\xwidth=#1pt \xshift=\hsize \advance\xshift
by-\xwidth \divide\xshift by 2
\yshift=#2pt \divide\yshift by 2
{\hglue\xshift \vbox to #2pt{\vfil
#3 \includegraphics{#4.eps}
}\hfill\raise\yshift\hbox{#5}}}

\def\lis#1{{\overline#1}}

\def\tende#1{\,\vtop{\ialign{##\crcr\rightarrowfill\crcr
 \noalign{\kern-1pt\nointerlineskip} \hskip3.pt${\scriptstyle
   #1}$\hskip3.pt\crcr}}\,}
\def\eg{{\it e.g.\ }}

\def\0{\noindent}
\def\*{\vskip2mm}
\def\media#1{\langle #1 \rangle}

\font\titolo=cmbx12%
\def\iniz{\setcounter{equation}{0}}
\def\be{\begin{equation}}\def\ee{\end{equation}}

\usepackage{fancyhdr}
\def\alert#1{{\color{ired}#1}}
\definecolor{iblue}{RGB}{65,105,225}
\definecolor{ired}{RGB}{220,20,60}
\definecolor{igreen}{RGB}{50,205,50}
\definecolor{ipurple}{RGB}{75,0,130}
\definecolor{iochre}{RGB}{218,165,32}
\definecolor{iteal}{RGB}{51,204,204}
\definecolor{imauve}{RGB}{204,51,153}

\iniz

\font\tengr=grreg10

\def\ifnextchar#1#2#3{\let\tempe #1\def\tempa{#2}\def\tempb{#3}\futurelet
  \tempc\ifnch}
\def\ifnch{\ifx\tempc\tempe\let\tempd\tempa\else\let\tempd\tempb\fi\tempd}
\def\gobble#1{}
\font\tengr=grreg10
\font\tengrbf=grbld10
\def\greekmode{%
\catcode`\<=13
\catcode`\>=13
\catcode`\'=11
\catcode`\`=11
\catcode`\~=11
\catcode`\"=11
\catcode`\|=11
\lccode`\<=`\<%
\lccode`\>=`\>%
\lccode`\'=`\'%
\lccode`\`=`\`%
\lccode`\~=`\~%
\lccode`\"=`\"%
\lccode`\|=`\|%
\tengr\def\bf{\tengrbf}
}
\newcount\vwl
\newcount\acct
\def\lt{<}

{
  \greekmode
  \gdef>{\ifnextchar `{\expandafter\smoothgrave\gobble}{\char\lq\>}}
  \gdef<{\ifnextchar `{\expandafter\roughgrave\gobble}{\char\lq\<}}
  \gdef\smoothgrave#1{\acct=\rq137 \vwl=\lq#1 \dobreathinggrave}
  \gdef\roughgrave#1{\acct=\rq103 \vwl=\lq#1 \dobreathinggrave}
  \gdef\dobreathinggrave{\ifnum\vwl\lt\rq140	
    \char\the\acct\char\the\vwl\else\expandafter\testiota\fi}
  \gdef\testiota{\ifnextchar |{\addiota\doaccent\gobble}{\doaccent}}
  \gdef\addiota{\ifnum\vwl=\lq a\vwl=\rq370
    \else\ifnum\vwl=\lq h\vwl=\rq371 \else\vwl=\rq372 \fi\fi}
  \gdef\doaccent{\accent\the\acct \char\the\vwl\relax}
}

\newif\ifgreek\greekfalse

\def\begingreek{\bgroup\greektrue\greekmode}
\def\endgreek{\egroup}

\let\math=$
{\catcode`\$=13
}

\def\begingreek{\bgroup\greektrue\greekmode}
\def\endgreek{\egroup}
\def\bgr{\begingreek}
\def\egr{\endgreek}

\begin{document}

\let\titolo=\bf
\alert{\centerline{\bf \Large Nonequilibrium Thermodynamics}}\vskip1mm

\centerline{\bf Giovanni Gallavotti} \centerline{\today}

{\vskip3mm}
\begin{abstract}: {\it A path from equilibrium to nonequilibrium thermodynamics}
\end{abstract}
          {\vskip3mm}
\def\SEC{Equilibrium}
\section{\SEC}
\label{sec1}

{\it Personne n'ignore que la chaleur peut \^etre la cause du
mouvement, qu'elle poss\`ede m\^eme une grande puissance motrice: les
machines \`a vapeur, aujourd'hui si r\'epandues, en sont une preuve
parlant \`a tous les yeux.}
\*

Equilibrium thermodynamics was born in the early 1800's. On the basis of
the experimental results on the properties of rarefied gases (like the
Boyle-Mariotte law) Carnot develops, \cite{Ca824}, his theorem on the ideal
efficiency of machines operating by extracting or damping heat in two
reservoirs and converting it into work. His long, very careful and
detailed, analysis is presented after a brief introduction (10 pages),
starting with the above words, in which present and future usefulness and
importance of the vapor machines is enthusiastically (and optimistically)
outlined (1824).

The theorem shows that the most efficient machines must operate running in
a reversible cycle in which the vapor (of water, air, alcohol or other
gases) evolves through a sequence $\PP$ of equilibrium
states;\footnote{Reversible transformations were essential in Carnot's
analysis, and he carefully insists to make clear the subtle argument that
permits to avoid regarding their definition, requiring for instance a
difference in temperature which ``can be considered as vanishing'', an
oxymoron: ``\`A la v\'erit\'e, les choses ne peuvent pas se passer
rigoureusement comme nons l'avons suppos\'e ..., \cite[p.13-14]{Ca824}.}
and this remains true even if the vapor is replaced by a liquid or a solid.

It is an example of what today we call a ``universal law'', \ie a law that
applies to a very large class of systems isolating, among their properties, a
few that they all verify in a quantitative form {\it without adjustable
  parameters}.

An immediate consequence is the possibility of defining the absolute
temperature of a heat reservoir: it is simply defined in terms of the
maximum efficiency of a machine operating between the reservoir of interest
and a fixed reservoir to which a conventional temperature value is
attributed.\footnote{\eg if the reservoir is water at its triple point then
in the Kelvin scale the absolute temperature is fixed to be
$T_0=273.16^oK$. The temperature $T_1$ of another reservoir at higher
temperature (say) is then given, in principle, by running a reversible
machine between the two reservoirs and deriving $T_1$ so that the
efficiency is $1-\frac{T_0}{T_1}$.  }

A few years later Kr\"onig, \cite{Kr856}, established the proportionality
of the absolute temperature to the average kinetic energy and Clausius,
\cite{Cl850}, wrote the first of the works leading to {\it entropy}, 1850,
whose existence constitutes the {\it second law} of thermodynamics and is
implied by Carnot's theorem.

The meaning of the word ``entropy'' was explained by Clausius himself
\cite[p.390]{Cl865}:\\``{\it I propose to name the quantity $S$ the entropy of
the system, after the Greek word \bgr<h trop'h \egr ``the transformation'',
{\rm [in German, Verwandlung]}. I have deliberately chosen the word entropy
to be as similar as possible to the word energy: the two quantities to be
named by these words are so closely related in physical significance that a
certain similarity in their names appears to be appropriate.}''

The notion of entropy became quickly fundamental for the theory and
applications of equilibrium thermodynamics and almost identified with it:
soon it was accompanied by the question of which would be its definition in
terms of the atomistic representation of matter.

And in 1866 Boltzmann, \cite{Bo866}, proposed to link it to the purely
mechanical Maupertuis's {\it Action Principle}, imagining that atoms were
moving on periodic orbits spanning all phase space points compatible with
the mechanical conservation laws.\footnote{The idea was again proposed four
  years later by Clausius, \cite{Cl871,Cl872}.}
 
This was a first attempt towards the formulation of the {\it ergodic
  hypothesis} and two years later, \cite{Bo868}, led to the description of
the statistical properties of a system in equilibrium via what are now
called {\it microanonical and canonical ensembles}. The consequences
developed by Boltzmann in 1868 were recognized by Maxwell (1879) who
commented them \cite{Ma879-c},\footnote{Interestingly {\it without even
    mentioning} the $H$-theorem, discovered in the meantime by Boltzmann.}
and by Gibbs, \cite{Gi902}.

Entropy $S_A$ is defined for every equilibrium state $A$ of a given system
and is useful to establish the balance between heat exchanged by a system
with the surrounding thermostats\footnote{\ie large bodies whose temperature and
volume variations are neglegible during the observation time} and work
performed on the surrounding mechanical devices.

A key property is that if a transformation through a ``path'' $\PP$ of
successive equilibrium states is {\it reversible} and leads
from an equilibrium state $A$ to an equilibrium state $B$ then the
the entropy variation is $S_B-S_A=\int_\PP \frac{dQ}T$,
if $dQ$ is the amount of heat that the system receives while in contact
with a heat reservoir at absolute temperature $T$, whatever the reversible
path $\PP$ is.  This allows to set the value of the entropy of a generic
equilibrium state, determining it up to an additive constant.

However entropy is important also in irreversible processes $\II$: it was
established by Clausius that in such a process $S_B-S_A\ge \int_\II
\frac{dQ}T$, where $T$ is the absolute temperature of the reservoir from which
the system receives the amount $dQ$ of heat.

In particular if the transformation $A\to A$ is along an irreversible path
$\II$ it is $\oint_\II\frac{dQ}T\le 0$.
Also if
$\II$ is an irreversible adiabatic path (\ie a sequence of transformations
with no heat exchange) it cannot lead from $A$ to $A$, and necessarily
leads from $A$ to a state $B$ with larger entropy.

If a system evolves first from a state $A$ to $B$ at constant temperature $T_2$
receiving a quantity of heat $Q_2$, then evolves
adiabatically from $B$ to $B'$ at temperature $T_1$, then to 
$A'$ at the temperature $T_1$ ejecting a quantity of heat $Q_1$ and
then adiabatically back to  $A$ it will be
$$\frac{Q_2}{T_2}-\frac{Q_1}{T_1}\le0,\quad {\rm
  which\ means:}\ \frac{Q_2-Q_1}{Q_2}\le \frac{T_2-T_1}{T_2}$$
Since $Q_2-Q_1$ must be, by energy conservation, the work performed by the
system in the cycle, the inequality means that the efficiency in the
transformation of heat into work cannot exceed $\frac{T_2-T_1}{T_2}$, and
it can equal it if the cycle is a reversible path: which is Carnot's
theorem formulated in terms of Clausius' entropy properties.

The continuous use of the second law, hence of Carnot's theorem, led to a
long debate on the ergodic hypothesis, on entropy, and on the resolution of
the antimony between reversible microscopic dynamics and irreversible
macroscopic evolutions, which in some respects is still going on.

\def\SEC{Microscopic stationarity}
\section{\SEC}
\label{sec2}

In recent times there has been a widespread interest in extending
Thermodynamics to a theory dealing with stationary states: these are states
of systems on which steady non-conservative forces may act, keeping
currents flowing through the systems: and currents can be of various kinds
like transporting matter, heat, electric charge, ...

The stationary states are a natural generalization of the equilibrium
states (which are very special cases of stationary states) and, of course,
a main question is whether general, system independent, properties can be
assigned to such states and be useful in studying their properties.

To understand the recent developments it is essential to keep in mind that
the study of a physical system starts from an initial datum $X$ where $X$
is a point in ``{\it phase space}'' $\FF$, often determined by the $6N$
coordinates of the $N$ molecules or atoms. And the point $X$ is generated
via suitable experimental devices (it has become common to say ``{\it
  following a prefixed protocol}'') and is a {\it random set of
  coordinates}, because of the many unavoidable uncontrolled actions
influencing the devices.

The probability of selecting the initial datum $X$ in $\FF$ is essentially
always unknown, but it is {\it assumed to have a density}: which means that
there is a function $\r(X)$ such that $\r(X)dX$ is the probability
that, repeating the experiment many times, the datum $X$ falls
in the volume element $dX$. Such a probability distribution is commonly
called ``absolutely continuous'' (with respect to the volume).

Here attention will be devoted to data $X$ generated randomly
with``absolutely continuous'' distribution. It will be seen that the
assumption of existence, behind the protocol of the observation, of an
unknown but absolutely continuous probability distribution for the initial
data generation is {\it far from obvious} and in the end it turns out to be
a {\it far reaching physical law}, that could be called law of ``initial
chaos''.

Given the initial data $X$, the system will evolve describing in time a
trajectory denoted $t\to S_t X=X(t)$.  In theoretical studies the evolution
$X(t)$ will be supposed to be defined by a solution $X(t)$ of a
differential equations $\dot X=F(X)$.

The interaction with the thermostats is influential and cannot
be ignored. This means that the evolution equations must involve the
interaction with the surroundings: which is a difficulty because it leads,
eventually, to consider infinite systems.\footnote{\ie interaction with the
  surroundings which interact also with their own surroundings, which
  interact with their own surroundings, ....}

The same difficulty is met even in equilibrium when it is imagined that the
temperature is fixed via the contact with a single thermostat.\footnote{In
  this respect Boltzmann in his 1868 paper shewed that in a very large system
  in a microcanonical equilibrium a finite volume subsystem behaves as if
  in contact with a reservoir with a temperature fixed and is described by
  a canonical ensemble.}

The difficulty is bypassed, in theoretical studies, by using dynamical
models involving finitely many particles obeying equations of motion in
which are introduced phenomenological forces (typically proportional to
currents, circulating in the system, via ``transport coefficients''). Such
forces model dissipative effects controlling the actions of the
non conservative forces or act by imposing suitable constraints for the same
purpose. This gives rise to models which involve forces that are believed
to be {\it physically equivalent} to more physical infinite thermostats but
contain {\it finitely many particles},
\cite{EM990,Ga013b}.

Given the equations of motion, an initial datum $X$ will evolve into $X(t)$,
reaching a {\it stationary state}.

Stationarity is in the sense that the observables will fluctuate in time
but will have well defined time independent ``{\it statistics}'', \ie
distribution of fluctuations around well defined time averages; and for
quite a few observables even their values will be essentially constant in
time (no need to average them) if the system is large (\ie it consists of a
large number of molecules).

In particular, from a microscopic viewpoint, the phase point representing
at time $t$ the microscopic state $S_tX$ of the system started in the
configuration $X$, will spend a fraction of the total observation time
$\th$ visiting, in the limit $\th\to\infty$, any chosen open
set\footnote{Hence any closed, or just ``measurable'', set in $\FF$.} $\DD$
with a well defined frequency $P_0(\DD)$.

Therefore imagining phase space divided into small regions
$\{\DD_i\}_{i=1,\ldots,\NN}$ the frequencies of visit $P_0(\DD_i)$ will
determine the average values of all observables $O$ whose variation in each
of the $\DD_i$ is negligible, so that their values can be denoted
$O(\DD_i)$: their time average will be $\media{O}=\sum_i
O(\DD_i) P_0(\DD_i)$.

Of course the size of the sets $\DD_i$ must be adapted to the observables
that are studied: therefore the choice of the $\DD_i$ can be regarded as a
choice of a ``coarse graining'' of phase space. 

The frequencies $P_0(\DD_i)$ will arise from a probability distribution
$P_0$ on $\FF$ {\it invariant} with respect to the evolution: in the sense
that if $X$ evolves in time $\t$ into $S_\t X$ then {\it any}
(reasonable, \eg open) set $\DD$ evolves into some $S_\t\DD$ and the two
sets are visited with the {\it same frequency} $P_0(\DD)=P_0(S_\t\DD)$.

A fundamental theorem, by Sinai in its simplest form, states that if
the motion of the system is ``{\it chaotic}'', as is the case for virtually all
systems ultimately constituted by atoms, then the probability is $1$ that a
protocol, of the above mentioned kind for obtaining the initial datum $X$,
generates a motion $S_tX$ which visits the sets $\DD$ in phase space with
well defined frequency $P_0(\DD)$, \cite{Si968a,Si968b,Si977,Bo975,BR975}.%
\*

\0{\it Theorem:} {\it there is a unique probability distribution $P_0$
  (called ``SRB distribution'') determining the frequencies $P_0(\DD)$ and
  therefore the statistical properties of the motion. Furthermore:\\
  (a) the
  frequencies $P_0(\DD)$ are independent on the protocol generating $X$,
  provided it generates random $X$'s with an absolutely continuous
  distribution,\\
  (b) and given the protocol the frequencies are
  $X$-independent with probability $1$,
  \\
  (c) and if there is one invariant
  distribution which is absolutely continuous then it must be $P_0$}.  \*

The requirement, aside from the mentioned law of initial chaos, is
that the equations of motion generate {\it chaotic motions in a
  mathematically precise sense} that will be briefly called ``{\it chaotic
  hypothesis}'', or CH, essentially meaning that
\*

\0{\it Chaotic Hypothesis: any datum $X$ evolves
  into $S_tX\equiv X(t)$ for $t>0$,
  \\
  (a) never stopping (\ie $\min |\dot
  X|>0$) and,
  \\
  (b) after a transient time $t_0$, an observer following
  $S_tX=X(t)$ and oriented as $\dot X(t)$
  will see $X(t)$ as a hyperbolic fixed point, \ie as a saddle point,
  and
  \\
  (c) data $X\in\AA$ evolve so that $S_tX$ covers densely a smooth
  surface $\AA$ (``attracting set'').}
\*

The theorem applies to any system with this property, whether isolated (as
in studying equilibrium) or in contact with external thermostats and under
action of non conservative forces (as in general stationary
states).\footnote{The requirement that the attracting set is a smooth
  surface can be weakened replacing ``smooth'' with ``closed'': in the first
  case the motion on $\AA$ is a ``Anosov system'' while in the second it
  satisfies ``Axiom A'', \cite{Ru989}.}

The result eliminates an essential difficulty. Leaving aside the need to
adapt the cells $\DD_i$ to the observables selected for analysis, if the
equations of motion generate chaotic motions\footnote{As it is the case
  apart from very few remarkable exceptions, like the arrays of elastic
  oscillators with interactions linear in the relative distances.} then
there are, as a mathematical theorem, many ({\it actually infinitely many})
stationary probability distributions $P\ne P_0$ that are invariant: \ie in
the cases covered by the above theorem there are data which visit the sets
$\DD$ with frequencies $P(\DD)$ {\it different} from $P_0(\DD)$, \ie
different from the SRB distribution.

Hence, without applying the theorem, requiring stationarity does not
provide a recipe to determine the statistical properties of the
motion.\footnote{In equilibrium the ergodic hypothesis, for instance, has
to be added to select the microcanonical ensemble distributions as the ones
describing the statistics of the motions.}

The theorem allows to select unambiguously {\it which is the distribution}
that determines the statistics $P_0$ observed in a given case, hence which
is the {\it physically important statistics} associated with the equations
of motion: it is valid both in equilibrium cases and in nonequilibrium 
ones.

Of course this implies accepting as a {\it physical law} that any protocol
generating the system initial configurations $X$ is ``absolutely
continuous'' and, furthermore, the evolution of such $X$'s is chaotic
(leaving aside evident exceptions): this law (proposed by
Ruelle, \cite{Ru989,Ru989b}, together with its main implications), cannot
be proved and it has to be accepted as a law of Physics.

The (idealized)\footnote{Because it is only possible to contemplate a
  protocol that generates initial data constrained to have a prefixed energy}
case of isolated systems, \ie the equilibrium cases for systems obeying
Hamiltonian equations of motion with a fixed energy, have the property that
one, $P_0$, {\it among the many stationary distributions} compatible with
the constraints (usually just with a fixed value of the energy), is such
that $P_0(\DD)$ can be expressed simply as an integral over phase space
$P_0(\DD)\equiv \int_\DD \r_0(X) dX$ of a density function $\r_0(X)$, which
exists as a consequence of Liouville's theorem.\footnote{Which states that
  any Hamiltonian system admits an invariant distribution with density over
  the surface of fixed energy (which is usually the only constraint).}

Hence $P_0$, under the chaotic hypothesis, determines the statistical
properties of the motions $X$.  It can be said that isolated systems not
only conserve the total energy but also admit an {\it invariant} absolutely
continuous way of measuring the phase space elements of the energy surface
(here invariant means that $P_0(\DD)=P_0(S_t\DD)$ for all $t,\DD$). This
also shows that the CH implies, for isolated systems, the ergodic hypothesis.

\def\SEC{Discrete representation of motion. Equilibrium.}
\section{\SEC}
\label{sec3}

Accepting the chaotic hypothesis, the problem of identifying the
probability distribution controlling the statistics of the observations, is
solved in general by the theorem in Sec.\ref{sec2}, for stationary states
of systems, whether {\it in equilibrium or not}.  And the question of how
to {\it extend thermodynamics to nonequilibrium stationary states} can be
posed, and one among the first questions is whether an extension of the
entropy as a state function is possible.

It is important to review first how the microscopic interpretation of entropy
arises in equilibrium thermodynamics. 

Via the $H$-theorem of Boltzmann (1872), the entropy of an isolated
rarefied gas is identified to be proportional to a quantity called $H$,
\cite{Bo872}.  Later (1884) Boltzmann proposed, {\it for general}
equilibrium cases (including liquids and solid materials), the entropy $H$
to be proportional to the logarithm of the phase space volume $W$ available
to the $N$ molecules (\ie the phase space points $X$ with total energy $E$
and with positions in a fixed volume $V$), namely: $S=k_B \log W$ with
$k_B$ being the Boltzmann's constant. This formula is
also consistent with the arbitrariness of the additive constant in the
entropy definition: the latter simply reflects the
arbitrariness of the unit employed to measure the volume $W$, which has the
dimension of the $3N$-th power of an action.\footnote{This is the form given to
  entropy by Planck: it is however implicit in several
  works of Boltzmann, among which \cite{Bo871-c,Bo884}.} 

At this point it is convenient to go back to the ergodic hypothesis in its
original formulation and see whether it can be applied also to stationary
nonequilibria and to the extension of thermodynamics to general stationary
states.

For some time it was apparently believed that, if the motion was so chaotic
that the trajectory $S_t X$ would become dense in the region of phase space
compatible with the constraints, then an invariant distribution would
necessarily have the above absolute continuity property, hence it would
have the form $\r_0(X)dX$ suggested by Liouville's theorem.

This seems to have been the opinion of Boltzmann and Maxwell and in support
of the latter statement the {\it ergodic hypothesis} was proposed. In the
words of Maxwell,\cite{Ma879-c}, on Boltzmann's work, \cite{Bo868}, where
the canonical and microcanonical ensembles were introduced it can be read:
``{\it The only assumption which is necessary for the direct proof {\rm [of
      what are now called the microcanonical and canonical distributions]}
  is that the system, left to itself in its actual state of motion, will,
  sooner or later, pass through every phase which is consistent with the
  equation of energy}''.

As the above quotation says, a phase space point $X$ evolves in time
visiting all $X'$ compatible with the constraints (in isolated systems
this meant all $X'$ with the same energy as $X$ and with
particles all located in their container $V$).

This can be consistent only if the phase space $\FF$ is not a continuum but
consists of a finite number of points on a regular lattice. The evolution
{\it in a fixed time step} $t$, small with respect to the duration of
atomic collisions, can be seen as a map $X\to X'$ in which the $X$
describing initially the system becomes after time $t$ another phase space
point $X'$. The map should be thought as a discretized solution of the
equations of motion (now {\it common in computer simulations}).  Thus the
evolution map $X\to X'$ having to visit all possible configurations is just
a {\it one cycle permutation} of the finite (very large) number of possible
$X$'s.

The ergodic hypothesis simply means that the permutation must be {\it
  cyclic}; the motions will be periodic and the uniformity of the lattice
forming the discrete phase space points on the energy surface implies that
the frequency of visit to a set $\DD\subset \FF$ equals the fraction
$P_0(\DD)$ of points contained in $\DD$ (if $\DD$, although small, contains a
very large number of discrete points). Hence average values of 
observables can be computed via a distribution on the energy surface which
is uniform (with respect to the surface elements area), \ie a
microcanonical distribution.

The assumption admits obvious exceptions (\eg harmonic lattices and integrable
systems) but it was believed to be quite generally correct, in the
discretized version of phase space, for isolated systems.

Maxwell, as well as Boltzmann and Clausius, worried about the boldness of
the assumption in the above form. About it Maxwell writes, \cite{Ma879-c},
``{\it But if we suppose that the material particles, or some of them,
  occasionally encounter a fixed obstacle such as the sides of a vessel
  containing the particles, then, except for special forms of the surface
  of this obstacle, each encounter will introduce a disturbance into the
  motion of the system, so that it will pass from one undisturbed path into
  another. The two paths must both satisfy the equation of energy, and they
  must intersect each other in the phase for which the conditions of
  encounter with the fixed obstacle are satisfied, but they are not subject
  to the equations of momentum. It is difficult in a case of such extreme
  complexity to arrive at a thoroughly satisfactory conclusion, but we may
  with considerable confidence assert that except for particular forms of
  the surface of the fixed obstacle, the system will sooner or later, after
  a sufficient number of encounters, pass through every phase consistent
    with the equation of energy}''.  

  The obvious objection is that the time scale for a phase space point $X$
  to go through a full orbit, \ie to visit all possible phase space points,
  must be, even for a small system (Boltzmann considered the example of
  $1\,{\rm cm}^3$ of hydrogen at normal conditions), unimaginably long:
  therefore the {\it ergodic hypothesis cannot be directly physically
    relevant} and has to be accompanied by proving other properties which
  explain why the equilibrium statistics can be observed within ``human
  time scales''.

Furthermore, even if the phase space points are supposed on a regular
lattice, it is not determined which would be their spacings. Their number
falling in a region $\DD$ remains determined up to a factor depending on
the lattice meshes $\d p$ and $\d q$ in the momenta and positions, \ie on
the precision with which the lattice represents the continuum: the number
of points in $\DD$ will be $\frac{vol(\DD)}{h^{3N}}$, where $h=\d p\d q$ is
a unit of action determined by the precision of the discrete representation
of the continuum.

Maxwell and Boltzmann addressed the time scale question, in the case of
rarefied gases, developing the Boltzmann's equation, which is an
approximation to the evolution, describing properties of a very restricted
class of observables. Nevertheless the class is wide enough to contain all
the observables obviously interesting for the theory of gas motions.

It leads to estimate the time scale necessary for a gas to relax to
equilibrium (described by the microcanonical distribution, as it would be
formally predicted by the ergodic hypothesis), or more precisely for
achieving a state in which the few observables that possibly are of
interest (like density, pressure, temperature, few particles correlations)
show a well defined value with relatively small fluctuations.

In other words, the {\it Founding Fathers} were well aware that the
ergodicity assumption could not be directly used to justify the approach to
equilibrium, as well as the time scale necessary for attaining the
equilibrium distribution.

They used it as a kind of 'symmetry property' that  {\it a priori} implied a
description of the equilibrium states in terms of the equilibrium
ensembles; however they insisted that what was really interesting and
physically necessary was that, for
each observable of interest, like the occupation numbers of domains
(``cells'') $\D$, in the $6$-dimensional single atom phase space, large
enough to contain a sizable fraction of the total number $N$ of atoms,
there should be a measurable time scale for reaching the average value.

The analysis can be found already in their attempts, by Maxwell, Boltzmann,
Thomson, to obtain macroscopic equations describing the evolution (to
equilibrium or stationarity) of several observables
(\cite{Ma867-b,Bo872,Th874b,Bo896a}).

\def\SEC{Entropy and nonequilibrium}
\section{\SEC}
\label{sec4}

Turning to nonequilibrium, the simplest case to keep in mind is that of a
gas in contact with two reservoirs at {\it different temperatures}. One can
also think to electrically charged particles moving in a lattice (with
periodic boundary conditions, modeling a ``wire'') of obstacles (molecules
of a ``crystal'') and subject to an electric field and to a thermostat. Or
to a fluid in a periodic container subject to a stirring force and in
contact with a thermostat to dissipate friction heat.

The equations of motion $\dot X=F(X)$ may have, and do have for several
models of nonequilibrium systems, a ``{\it time reversal symmetry $I$}'':
here $I$ is a smooth map of phase space and the solution $S_tX$ with
initial datum $X$ has the property that $IS_t X=S_{-t}IX$.

Can one proceed, as done in equilibrium, and imagine the phase space
configurations $X$ compatible with the constraints as a discrete set of
points located in the usual continuum phase space? This is tempting as it
would bring back the idea that the motion of a phase space point wonders
visiting successively all other points: it would also explain the existence
of a {\it unique stationary distribution}, which would be simply the
distribution giving equal weight to all points.

It would be natural to form a partition $\PP$ of the continuum phase space
of a system into finitely many sets $\DD_i$ and call $P_0(\DD)$ the
probability attributed to each set $\DD$ by the invariant SRB distribution,
\ie the frequency of visit to $\DD$ from an initial datum $X$ obtained
with probability $1$ via some protocol (see Sec.\ref{sec2}). Such
probability is well defined {\it although generally not expressible} by an
integral over $\DD$. And then {\it replace} the continuum phase space by a
{\it finite number} $\NN$ of points, with $\NN P_0(\DD)\gg 1$ of them in
each $\DD\in\PP$.

Furthermore the evolution should be a {\it one cycle permutation} of the
phase space points: in this way each cell $\DD$ is visited (in a very long
time) with a frequency $P_0(\DD)$ which is, therefore, uniquely determined
and is a representation of the SRB distribution. The time necessary needs
not be too long if the cells $\DD$ are not too small (although small enough
so that the observables of thermodynamic interest can be considered
constant in each of them) and contain a fraction of the total number of
discrete points of order $1$ (so that $\NN P_0(\DD)\gg 1$). 

But in the case of nonequilibrium the equations of motion are no longer
Hamiltonian and are dissipative. This is manifested by the {\it divergence}
$\s(X)$ of the equations of motion:\footnote{Given a general ODE $\dot
x_j=f_j(x), j=1,\ldots,n$ the divergence definition is
$\s(x)=-\sum_j \dpr_{x_j} f_j(x)$ and gives the rate of
compression of a volume element $dx$ around $x$; it can be $>0$
(compression) or $<0$ (expansion) depending on $x$.}  which is not $0$ (as
it is for the isolated evolutions, \ie in the Hamiltonian cases) and must
have a $\ge0$ average $\media\s$.\footnote{The average $\media\s$ cannot
be $<0$, \ie phase space cannot keep expanding forever while a stationary
state is reached.}

If the ``chaotic hypothesis'' (CH) holds, $X(t)$ evolves towards a surface
$\AA$\footnote{Possibly of dimension lower than that of phase space if the
forces that keep the system out of equilibrium are not small enough.}  and
initial data $X$ starting out of it evolve in time with their distance to
$\AA$ tending to $0$ (exponentially fast). The surface $\AA$ can possibly
be different from the entire phase space compatible with the constraints
and have lower dimensionality: in any case if $\media{\s}>0$ the statistics
will be a probability distribution which gives probability $1$ to a subset
in $\AA$, the ``{\it attractor}'' $\AA_o\subset\AA$, which has $0$ volume,
or $0$ surface area if $\AA$ is a surface of lower dimension.

A discretization of phase space should therefore be a discrete
representation of the attractor $\AA_o\subset\AA$ that can be imagined
replacing the continuum phase space by a {\it regular} lattice in which the
position and momentum coordinates are on a grid spaced by $\d q$ and $\d
p$, and then {\it discard} the points that are not on the cyclic
permutation toward which data generated by the initial protocol
evolve.\footnote{The regularity of the lattice representing the discretized
phase space reflects the special relation (called absolute continuity in
Sec.1) between the protocols that generate initial data and the volume
measure. In the equilibrium cases (\ie Hamiltonian evolutions) all grid
points are thought to be part of the evolution cycle, as a literal
interpretation of the ergodic hypothesis.}

Under the CH {\it heuristic arguments} can be developed to {\it estimate
the number $\NN$ of discrete points} necessary to give an {\it accurate
description} of the motions of data on the attractor
$\AA_0\subset \AA$, \cite[Sec 3.11]{Ga013b}, out of the $\NN_0$ regular
grid points of the discretized phase space. Once the discretization is
obtained and $\NN$ is estimated, it is will tempting to define entropy as
proportional to $\log\NN$.

However the result is that $\log\NN$ might {\it not be defined up to an
additive constant} depending only on the precision of the discretization:
but changing the precision (\ie the size of the discretization meshes) it
changes by a quantity which depends {\it also} on the stationary state
considered, in particular it depends on the average phase space
contraction $\media{\s}$: this is in sharp contrast with the
equilibrium result where changing the precision changes $\log\NN$ by a
constant {\it independent} of the equilibrium state studied.
  
This indicates\footnote{But does not ``prove'', even under the CH, because the
estimates are heuristic.} that entropy, as a {\it function of state}, might
not be definable for stationary states out of equilibrium,
\cite[Sec.3.10,3.11]{Ga013b}, unless a physical constraint determining the
maximum precision of the discretization can be found (but such quantization
of phase space would require extra information).

However one of the main features of the {\it extension} of entropy to
rarefied gases not in equilibrium, but isolated and evolving towards
equilibrium, is that it is a ``Lyapunov function'' varying with time and
approaching (monotonically) a maximum value as a limit value, namely the
equilibrium entropy, \cite{GGL005,GHLS018}.

It is conceivable that in the evolution to a stationary state it could be
possible to define a Lyapunov function with the same property of evolving
(possibly not monotonically) to a maximum which is reached at stationarity,
\cite{Ga001,Ga013b}. 

If an initial non stationary distribution is considered (including possibly
a distribution specifying a single phase space point $\x_0$) then the
fraction $P(\x,t)$ of times in $[0,t]$ that the point $S_t\x_0$ visits $\x$
tends to $\frac1\NN$, as prescribed by the SRB distribution in the above
discrete representation, where $\NN$ is the number of points in the discretized
attractor $\AA_0$. Then $S(t)=k_B\sum_\x
  -\lis{P(\x,t)}\log\lis{P(\x,t)}$ tends, as $t\to\infty$, to:
$$S_\infty=k_B\sum_\x -\frac1\NN \log \frac1\NN=k_B\log\NN,$$
hence $S_\infty$ is the maximum value that $S(t)$ can
reach\footnote{because the maximum of $-\sum_{i=1}^M p_i\log p_i$ is $\log
M$ and is achieved when $p_i=M^{-1}$.} and therefore $S(t)$ can play the
role of a Lyapunov function.

The function $S_{\infty}$ depends {\it non trivially} on the precision of
the discretisation (as just discussed); and changing just the
precision, \ie the discretization mesh, $S_\infty$ will change depending
nontrivially on the particular stationary state.  So, when studying the
stationary states of a system, as functions of the parameters entering into
its equations of motion, it does not seem that $S_{\infty}$ can defined
just up to an additive constant independent of the particular stationary
state, hence {\it cannot be considered a function of state}.

The question on whether it is possible to define a function of state
generalizing the entropy function to nonequilibrium steady states remains
an interesting open question.

Still for all choices of the discretisation $S(t)$ will have the property
that it reaches the maximum value on the SRB distribution, \ie on the
natural stationary state. Entropy, as a function of state, may not defined
in general stationary states although the approach to stationarity may
admit a Lyapunov function (possibly related to the above $S(t)$): the
latter would extend the role plaid in the approach to equilibrium by the
entropy function as recently defined beyond the rarefied gases by the wider
interpretation of the formula $S=k_B\log W$, \cite{GGL005,GHLS018}.

\def\SEC{Quest for universality}
\section{\SEC}
\label{sec5}
\*

The remarkable general validity of the distributions giving the statistical
properties of the equilibrium states of very general systems is extended to
general stationary states via the associated SRB distributions, as described
by the theorem in Sec.\ref{sec2}. The latter however do not have the same
character as they may seem, at first, too theoretical.%
\footnote{As opposed to the equilibrium cases where the ergodic hypothesis
  yields a concrete universal prescription to determine the statistics of
  the motions to be given by the ``Gibbs states''.} However the problem may
  simply be due to the still recent introduction of the SRB
  distributions.\footnote{The complaints, that can sometimes be found in
  the literature, on the lack of explicit examples of many degrees of
  freedom systems are not well founded because examples which are chaotic,
  reversible (or not) and admit SRB distributions do
  exist, \cite{BK995,GBG004,BFG004,BGG007}.}

Universal relations very often reflect symmetry properties: the
Onsager reciprocity relations are a manifestation of the basic time
reversibility of the equations of motion and a first example of general and
universal property for nonequilibrium systems. However they are a property
that holds only as a first order approximation in terms of the size of the
active forces determining the nonequilibrium dynamics.

It is natural, as a first step in searching for universal properties, to
wonder whether the reciprocity relations have an extension to general non
equilibria, as a consequence of the time reversibility, that is {\it always
  valid} in the basic equations, even in presence of dissipation.

The question can be studied for systems in contact with thermostats and
subject to non conservative forces so that their stationary states (if
existing) will be out of equilibrium. Models of thermostats have been
introduced which involve forces that are believed to be equivalent to
purely mechanical infinite thermostats but {\it involve finitely many
  particles} and are described by {\it reversible equations},
\cite{EM990,Ga013b}.\footnote{Here, as in the previous section, time
  reversal is a {\it smooth map} of phase space such that $I S_t
  X=S_{-t}IX$.}

For such systems, particularly important in numerical simulations, time
reversal is a valid symmetry and a preliminary problem is to see how comes
that a {\it reversible equation leads to irreversible evolution}. 

Dissipation is controlled by the phase space contraction, \ie the
divergence $\s(X)$ of the equations of motion: and $\s(X)$ is a
particularly interesting observable (which often has the interpretation of
thermostats entropy increase rate). Time reversal symmetry implies that for
each configuration $X$ with $\s(X)>0$ there is another $IX$ with
$\s(IX)<0$: reversibility implies that dissipation (\ie phase space
contraction) has to coexist with phase space expansion.

In dissipative systems motions evolve towards an attracting set $\AA$ which
is not necessarily time reversal symmetric. Followed backwards in time they
also evolve towards an attracting set $\BB=I\AA$, which however attracts
points only in the backward evolution: and $\BB$ is a repelling sets for
the evolution forward in time if $\BB\ne\AA$.

However if the forces driving the system to a stationary nonequilibrium
state are small enough, although not infinitesimal as they are in the
theory of Onsager reciprocity, and if the ``chaotic hypothesis'' can be
assumed it follows that both $\AA$ and $\BB$ {\it are the same} because
motions with any initial data are dense on phase space.\footnote{This is
consequence of the ``structural stability'' property of the CH.}
Nevertheless the SRB distribution on $\AA=\BB$ for the forward motion and
the SRB distribution for the backward motion are {\it mutually singular}.

In other words there is a fractal set $\AA_+\subset\AA$ of data (a
``forward attractor'') which have probability $1$ to generate the SRB
distribution $\m_+$ for the forward motions and probability $0$ to generate
the SRB $\m_-$ for the backward motions, and vice-versa.\footnote{the
``attractor'' $\AA_+$ on $\AA$ is not uniquely defined: usually it is an
invariant set which has $\m_+$-probability $1$ and has maximal fractal
dimension: typically it can be modified by subtracting from it the
trajectory of one or more data $X$.}

Actually data in $\AA_+$ run forward or backward in
time generate the same SRB statistics $\m_+$ and the corresponding
statement holds for the SRB statistics $\m_-$: it is $\m_\pm(\AA_\pm)=1,
\m_\pm(\AA_\mp)=0$. So the irreversibility is made manifest, in such time
reversible systems, by the different statistics obeyed with probability $1$
by motions generated by a protocol as discussed in Sec.1 and observed
as time tends to $+\infty$ or to $-\infty$.

Therefore it is interesting to study the effect of time reversal symmetry
on systems in which the attracting set $\AA$ is equal to its time reversal
$I \AA$ and both coincide with the phase space available (although the
forward and backward SRB statistics are mutually singular). In general such
motions {\it taking place on} $\AA$ will have a surface contraction rate $-\s$
with a {\it positive time average} $\media{\s}>0$ of $\s$, in the forward
evolution ({\it as well as} in the backward evolution).

For such systems the fluctuations of the dissipation $\s(S_t X)$, as
the initial data $X$ evolve into $S_tX$ and $t\to+\infty$, have remarkable
properties because, under the chaotic hypothesis, the motions on $\AA$ are
in several aspects are well understood.\footnote{Under the CH the evolution
on $\AA$ is called a ``Anosov system'': such systems can be considered as
the paradigm of chaotic motions, and in the theory of chaotic dynamics
play a role similar to that payed by harmonic oscillators for regular dynamics.}

Define the function: ``probability $P(p,\t,\d)$ of the set of $X$ such that
$\frac1\t\int_0^\t \frac{\s(S_tX)}{\media{\s}}dt\in [p,p+\d]$'': this
function is interesting because often $\s(X)$ has the physical
interpretation of {\it entropy generated} (in the thermostats) per unit time by
the stationary state considered.

Then a general theorem applies to the motions on $\AA$, essentially
comparing the probability that $p$ has a given value to that of having the
opposite value, \ie the probability of entropy production in time $\t$
equal to $\sim p\media{\sigma} \t$ to that of $\sim -p\media{\sigma} \t$:
{\it if CH holds and the motions are time reversible and their trajectories
are dense on phase space\footnote{for instance, under the CH, at small
forcing of a thermostatted Hamiltonian system.}} the probabilities can be
shown to be expressed, to leading order for large $\t$, via a density $\sim
e^{ s(p)\t}$ and :\*

\0{\it Fluctuation Theorem:\kern3mm $s(-p)=s(p)-p\,\media{\s}\,\t$}
\*

\0with $s(p)$ convex and maximal at $p=1$:  which is a {\it universal} relation
in the sense that it contains no free
parameters, \cite{GC995,Ge998}.
 
It is remarkable, in particular, that it establishes a relation between the
``normal'', \ie most probable, value of the average entropy production in a
time $\t$, corresponding to $p=1$, to the highly unlikely ``opposite event''
with $p=-1$, and the relative probability depends on the entropy production
rate $\media\s$ but is otherwise independent on the system considered.

The above relation is called {\it fluctuation theorem}, FT, when the
mathematical assumptions are satisfied, or {\it
fluctuation relation}, FR, if the hypotheses are considered
phenomenologically valid. There are few examples in which time
reversibility holds, the models have a nonequilibrium stationary state, and
the FR holds, although the FT hypotheses, would be to difficult to check
(if at all holding), \cite{DLS002,GDL010}.

It shows that in such systems the entropy of the thermostats grows by
$\sim\media{\s} \t$ in time $\t$ with a probability $e^{\media{\s}\t}$ larger
than the probability that it decreases by the same amount, to leading order
in $\t\to\infty$.

It has been shown that, under the chaotic hypothesis, the fluctuation
theorem implies Onsager reciprocity so the FR provides an extension to
Onsager's theorem, \cite{Ga996,GR997}, at small but not necessarily
infinitesimal forcing.

The fluctuation relation is a first universal relation found, under the CH,
for general nonequilibrium systems at small forcing. And the question is
whether it is of any interest for systems which are not described by
reversible equations of motion: since most macroscopic equations involve
frictional forces the applicability of the FR may seem restricted, at most,
to microscopic and small systems. But time reversal is even ``behind''
macroscopic equations derived phenomenologically and containing forces
explicitly violating time reversal (like friction).

This suggests that the same phenomena could be equally described by
equations which are time reversible: such equations would show a variable
phase space contraction $\s(X)$ (unlike the  familiar time irreversible
ones, which typically show a constant phase space contraction, proportional
to some transport coefficients, \eg friction).  The divergence $\s(X)$
arising in the reversible models can be regarded as a special
observable. As such it can be studied in the model with irreversible
evolution $S^{irr}_tX$: if the two descriptions are equivalent it can be
expected that the fluctuations of $p=\frac1\t\int_0^\t \s(S^{irr}_t X)dt$
also satisfy the FR. If so the FR could be observable even in irreversible
models (if they derive from microscopic models which satisfies time
reversal, as they should), \cite{Ga018}.

The possibility of having the same system described by irreversible or
reversible equations indicates interesting analogies with the equivalence
of ensembles in equilibrium statistical mechanics: a subject to which some
attention is currently devoted, \cite{Ga018}.

\bibliographystyle{unsrt}

\begin{thebibliography}{10}
\bibitem{Cl850}
R.~Clausius.
\newblock On the motive power of heat, and on the laws which can be deduced
  from it for the theory of heat.
\newblock {\em Philosophical Magazine}, 2:1--102, 1851.
\end{thebibliography}

\end{document}